\newcommand{\be}{\begin{equation}}
\newcommand{\ee}{\end{equation}}
\documentstyle[aps,epsf,multicol,prl]{revtex}
\begin{document}
\draft
\widetext
\title{Probing the plateau-insulator quantum phase 
transition in the quantum Hall regime}

\author{R.T.F. van Schaijk and A. de Visser}

\address{Van der Waals-Zeeman Institute, University of Amsterdam, 
Valckenierstraat 65, 1018 XE Amsterdam, The Netherlands}

\author{S. Olsthoorn}

\address{High Field Magnet Laboratory, University of Nijmegen, Toernooiveld,
6525 ED Nijmegen, The Netherlands}

\author{H.P. Wei}

\address{Department of Physics, Swain Hall West, Indiana University, Bloomington, Indiana 47405}

\author{A.M.M. Pruisken}
\address{Institute for Theoretical Physics, University of California at Santa
Barbara, California 93106-4060}
\address{and}
\address{Institute for Theoretical Physics, University of Amsterdam,
Valckenierstraat 65, \\ 1018 XE Amsterdam, The Netherlands}

\maketitle

\begin{abstract}
\noindent We report quantum Hall experiments on the plateau-insulator
transition 
in a low mobility $In_{.53} Ga_{.47} As/InP$ heterostructure.
The data for the longitudinal resistance $\rho_{xx}$ follow an
exponential law and
we extract a critical exponent $\kappa \!=\! .55 \!\pm\! .05$ 
which is slightly
different from the established value 
$\kappa\! =\! .42 \!\pm\! .04$ for the plateau transitions. 
Upon correction for inhomogeneity
effects, which cause the critical conductance $\sigma_{xx}^*$ to
depend {\em marginally} on temperature, our data indicate that the plateau-plateau and plateau-insulator 
transitions are in the same universality class.

\end{abstract}


\begin{multicols}{2}
\narrowtext

In the field of two dimensional electron gases
the nature of the transitions between adjacent quantum Hall plateaus ($PP$
transition) is an
ardent topic of research. Experiments on 
low mobility $In_{.53} Ga_{.47} As/InP$ heterostructures are a remarkable
demonstration of 
a {\em quantum phase transition} indicating that the quantum Hall steps
become infinitely sharp as the temperature
($T$) approaches absolute zero.$^{1,2}$
The maximum slope in the Hall resistance ($\rho_{xy}$) with varying
magnetic field $B$
was shown to diverge algebraically in $T$, 
$({\rm d}\rho_{xy} /{\rm d}B)_{\rm max}
\!\propto\! T^{-\kappa}$ with the critical exponent
$\kappa\!=\! .42 \pm .04$. On the other hand,
the half width ($\Delta B$) of the longitudinal resistance ($\rho_{xx}$) 
was shown to vanish like  $\Delta B \!\propto\! T^{\kappa}$ with the same
exponent~$\kappa$.  

Due to the short range random alloy 
potential scattering, the low-mobility $In_x Ga_{1-x} As/InP$ 
structure has proven to be exceptionally important for studying scaling
phenomena. This produces a wide range in $T$ where the transport is
dominated by Anderson (de)localization
effects. This is in sharp contrast to 
high mobility $GaAs/Al_x Ga_{1-x} As$ heterostructures where the long-range
potential 
fluctuations dramatically complicate the observability of the critical
phenomenon, given the limitations of the
experiment.$^{3,4}$

Nevertheless, the $PP$ transitions in
$GaAs$ heterostructures have  been studied extensively. It is therefore not
surprising
that the {\em universality} of the plateau transitions has not been
established by these investigations.
In these experiments,
the same value of $\kappa \!\approx\! .42$ was found but for a few samples only
and for a small range in lowest $T$.$^4$
However, in most of the samples simple data fitting produced $\kappa$'s
ranging from
$.2$ up to $.9$. These results are Landau level dependent and even for a
given Landau level 
the $\rho_{xx}$ and $\rho_{xy}$ data give rise to different values for
$\kappa$.$^{5}$

The focus in the last few years has been on transport in the lowest Landau
level. 
For this purpose samples of lower density were used.$^{6-10}$ 
The bare resistance data look quite different from those of
the other Landau levels since the transition is now between a quantum Hall
(plateau) phase and an {\em insulator} ($PI$ transition). There are
striking similarities between the $PI$
transition at high $B$ and the superconductor-
insulator transition$^{11}$ and the metal-insulator transition in two
dimensions at $B \!=\! 0$.$^{12}$

One of the most important predictions of the renormalization theory is that
the $PP$ and $PI$
quantum phase transitions are the same.$^2$ This stipulates that the same
$\kappa$ be observed
as $T$ approaches absolute zero and 
that the (electron-hole) symmetry in the $\sigma_{xx}$, $\sigma_{xy}$
conductance plane be retained. 
These aspects of Landau level systems serve as an important guide for
selecting samples, particularly
since the range of available $T$ in transport measurements is rather
limited and usually not sufficient to establish 
a critical point. 
However, in the experiments of Refs 6-10 a comparison between the $PP$ and $PI$
transitions within the same sample was either not possible or not drawn. 
Moreover, these experiments provide inconclusive results because the
samples used suffer from the aforementioned complications. 

Recently, an interesting empirical result for the lowest Landau level
$\rho_{xx}$ has been reported.$^{13}$
For arbitrary samples at finite $T$, the $\rho_{xx}$ data seems to depend
linearly rather than algebraically
on $T$, indicating that the problem is generically the same for all $GaAs$
samples. Once again, the experimental design 
has overlooked an essential requirement for studying scaling phenomena:
the importance of short range random potential scattering - an essential
prerequisite for sample choice.

In this Letter we report the results of magneto transport experiments
performed  
on an $In_{.53} Ga_{.47} As/InP$ heterostructure. Our main objective 
is to study critical aspects of the $PI$ transition and to compare the
results to 
the $PP$ transitions measured on the same sample. 
We benefit from the fact that our sample has been studied before.$^{14}$ 
In particular, the exponent $\kappa$ for the $PP$ transitions was found to
be $.42$ and $.20$ for spin polarized and
spin degenerate Landau levels respectively. The mobility of the sample is
$\mu  \!=\!  16000 cm^2 /Vs$ measured at $T \!=\!  4.2 K$.
The electron density is $2.2 \cdot 10^{11} cm^{-2}$ which means that the
$PI$ transition occurs at $B \!\approx\! 16 T$. 

Our experiments were carried out in a Bitter magnet ($B \!<\!  20T$) using a 
plastic dilution refrigerator 
($.1 \!-\! 2 K$) and a bath cryostat ($1.5 \!-\! 4.2K$).
The magneto transport 
properties were measured using a standard ac-technique with a frequency of
6Hz and an excitation current of 5nA.
The main experimental results are presented in Fig. 1, where we have traced
$\rho_{xx}$ and $\rho_{xy}$ as $B$ 
sweeps through the $PI$ transition. The different sets of data show a
characteristic value $B \!\approx\! 16T$ 
separating the insulating phase at high $B$ and the quantum Hall phase at
lower $B$. 
The $\rho_{xy}$ data are shown in the inset of Fig. 1. This quantity, at
low $T$, is clearly 
not quantized through the transition. 

In order to recognize the relevant structure in the combined 
$\rho_{xx} , \rho_{xy}$ data we have computed the conductance components
$\sigma_{xx}$ and $\sigma_{xy}$ 
in the standard fashion. The results for different $T$ are plotted 
in the $\sigma_{xx} , \sigma_{xy}$ flow diagram (Fig. 2 upper inset). 
The symmetry about the line $\sigma_{xy}  \!=\!  1/2$ is striking 
and reflects the high quality of the experimental data. 
The same symmetry was observed and discussed in the original work on the
$PP$ transitions.$^{15}$ 
Following the renormalization 
theory of the quantum Hall effect$^2$ we extract the critical value $B_c$
from the maximum in $\sigma_{xx}$ or from 
$\sigma_{xy}  \!=\!  1/2$. 
The $B_c$ ranges from $16.3T$ at $.13K$ up to $16.9T$
at $4.2K$, indicating a change in electron 
density of $3.5{\%}$. This is smaller than what is
obtained from the low $B$ Hall 
data, which gives a density of 
$ 2.0 \cdot 10^{11} cm^{-2} $ at $.13K$ as compared to $ 2.2 \cdot 10^{11}
cm^{-2}$ at $4.2K$. 
The symmetry
in the $\sigma_{xx}, \sigma_{xy}$ diagram is a direct consequence of the
following relations 
which hold at low but fixed $T$

\be
\sigma_{xx} (\Delta\nu) = \sigma_{xx} (-\Delta\nu)\; ; \; \sigma_{xy}
(\Delta\nu) = 1-\sigma_{xy}
(-\Delta\nu).
\ee
Here $\Delta \nu  \!=\!  {1\over B} \!-\!  {1\over {B_c}}  \!=\!  \nu  \!-\! \nu_c$.
We have explicitly verified the validity of Eq. 1. This result is important
since it 
fundamentally reflects the electron-hole symmetry in the problem. Our data
do not follow the 
statement of "duality"$^{16, 17}$ which says that 
$\rho_{xx} (\Delta \nu)  \!=\! 
\rho_{xx}^{-1} ( \!-\! \Delta \nu)$
and $\rho_{xy}$ remains quantized through the $PI$ transition. Instead, we
observe that the 
$\sigma_{xx}$ peak at $\Delta \nu  \!=\! 0$ develops a maximum around {\em 1K}.
Thus only for $T  \!<\!  1K$ the asymptotic scaling regime for the $PI$ transition 
is attained.$^1$ For the $PP$ transitions this range is larger ($T  \!<\!  3K$).

Since the $\sigma_{xx} , \sigma_{xy}$ data show the characteristic Landau
level independent behavior
we can now extract the critical exponents in a similar fashion as done
previously 
with the $\rho_{xx}$ and $\rho_{xy}$ data from the higher Landau levels. 
For example, from $\sigma_{xx}$ with varying $B$ we obtain the 
half width $\Delta B \!\propto\! T^\kappa$ with an exponent $\kappa  \!=\! .46 \pm
.05$. On the other hand we find 
$({{\rm d}\sigma_{xy}\over {\rm d}B})_{\rm min} \!\propto\! T^{-\kappa}$
with $\kappa  \!=\!  .43 \pm .05$. We attribute the small difference between the 
exponents to the uncertainties caused by mixing the $\rho_{xx}$, $\rho_{xy}$
data in the computation of the conductances.$^{15}$
In Fig. 3 we plot the $(\Delta B)^{-1}$ vs. $T$ for both the $PI$ and 
the $2\rightarrow 1$ plateau transition. The latter was derived from the half width in 
$\rho_{xx}$. For later purposes we have also plotted the low $T$ data for 
$({{\rm d}\sigma_{xy}\over {\rm d}B})_{\rm min}$ vs. 
$T$ in the lower inset of Fig. 4. The resulting exponents 
$\kappa  \!=\! .46 \pm .05, .42\pm .05$ and $.43\pm .05$ are all the same,
within the experimental error, indicating
that the $PP$ and the $PI$ transitions are in the same universality class.

Next we compare our data to the exponential result$^{13}$ 
$\rho_{xx}(\nu,T) \!\propto\! \exp( \!-\! \Delta\nu /\nu_0 (T))$. 
In Fig. 2 we have traced
$\rho_{xx}$ on a log scale as function of the difference $\Delta\nu$. 
This gives an adequate description of the data. The slope ($\nu_0^{-1}$) 
of the straight lines at the transition regime
can be accurately determined at each $T$. In Fig. 3 we plot $\nu_0^{-1}$
versus $T$ on 
a log-log scale. The data nicely follow the algebraic behavior 
$\nu_0^{-1} \!\propto\! T^{-{\kappa'}}$ with 
$\kappa'  \!=\!  .55 \pm .05 $. This value differs from 
the expected value $\kappa  \!=\!  .42 \pm .05$ by 
more than the experimental error. 

It is important to remark that the data do not follow the
linear behavior $\nu_0  \!=\!  \alpha T \!+\! \beta$ 
as proposed and advocated in Ref. 13. 
Such linear dependence on $T$ clearly does not describe the asymptotics of
the quantum phase transition
at $T$ equal to zero. Instead, it is semiclassical in nature and typically
observed at finite $T$ and on samples with 
predominantly slowly varying potential fluctuations.$^{3}$ 

In order to show that the value 
$\kappa'  \!=\!  .55 \!\pm\! .05$ is not a specific
property of the
$PI$ transition, we have mapped the
$2\rightarrow 1$ plateau transition onto the lowest Landau level 
following the steps $\rho_{xx} , \rho_{xy} \rightarrow \sigma_{xx} ,
\sigma_{xy} \rightarrow
\sigma_{xx} , \sigma_{xy}  \!-\!  1 \rightarrow 
\rho_{xx} , \rho_{xy}$.
The $\rho_{xx}$ data thus obtained were fitted to the exponential
expression leading to a value 
$\kappa'  \!=\!  .51 \!\pm\! .05$ 
(see Fig. 3). Transformations like this generally lead to less quality
data. Nevertheless the results 
in Fig. 3 indicate that different exponents can be extracted
from the same experimental data.

We next address the origin of this difference. First we point out that the
transport data of the $PI$ transition are 
accurately described by writing
\be
\rho_{xx} (\nu,T)= \rho^* (T)\exp( \!-\! T^{-\kappa'} \Delta\nu).
\ee
Here  $\rho^*$ denotes the {\em critical} resistance. It can be
written as  
$\rho^*  \!=\!  \sigma_{xx}^* /((\sigma_{xx}^* )^2+1/4)$ where $\sigma_{xx}^*$
stands for the peak in $\sigma_{xx}$. 
Both quantities are weakly dependent on $T$ and quite surprisingly, this
$T$ dependence 
is not simply {\em irrelevant} as thought previously. It is, in fact, {\em
marginal} and we next show that 
it accounts for the difference in the observed exponents.  

Following Eq. 1 we can relate $\rho_{xx} (\Delta\nu)$ to 
$\rho_{xx} ( \!-\! \Delta\nu)$ such that we can write the ratio as
$\rho_{xx} (\Delta\nu) / \rho_{xx} ( \!-\! \Delta\nu)  \!=\!  \exp( \!-\! 2T^{-\kappa'} \Delta\nu)$.
As a good check upon the validity of this result
we have fitted the exponential on the r.h.s. to the experimental data which
were inserted into  
the l.h.s. The same numerical value $\kappa'  \!=\!  .55 \pm .05$ was obtained
indicating once more that Eq. 1
represents the fundamental symmetry in the problem. From the ratio we obtain 
the following renormalization group equation for small 
$\theta \!=\! \sigma_{xy} \!-\! 1/2$
\be  
{{\rm d}\theta \over {\rm d}\ln T} = -\kappa \theta\;\;\;\;  ; \;\;\;\; 
\kappa = \kappa' - {{\rm d}\ln (\sigma_{xx}^{*2} + 1/4)\over {\rm d}\ln T}.
\ee
Eq. 3 shows how a relatively weak $T$ dependence in $\sigma_{xx}^*$ can
lead to different exponents 
extracted from different quantities.
In Fig. 4 we replot $1/\nu_0$ versus $T$ on a log-log scale. The solid
line gives 
$\kappa'  \!=\!  .55$. In the upper inset we plot the low $T$ data for
$\ln((\sigma_{xx}^* )^2+1/4)$ vs. $\ln T$ and obtain a 
slope of $.15 \pm .03$. According to Eq. 3 we have $\kappa \!\approx\!
.40$. This value should be compared  
with $\kappa  \!=\!  .43$ extracted 
from the low $T$ dependence of 
$({\rm d}\sigma_{xy} /{\rm d}B)_{\rm min}$ (Fig. 4). 

A naive interpretation of the {\em marginal} dependence of $\sigma_{xx}^*$
on $T$ is that the 
electron gas has not yet
fully developed criticality. This would mean that a much lower $T$ is
necessary before the 
critical fixed point is truly reached. 
However, it is important to stress that the    
small changes in $\sigma_{xx}^*$ observed at low $T$ are most likely the
result of macroscopic inhomogeneities
in the sample. One way of showing this is by writing Eq. 2 as

\be
\rho_{xx} (\Delta \nu, T) = e^{-{{\Delta\nu-\delta\nu_c (T)}\over{\nu_0 (T)}}}.
\ee
Here the shift in the critical filling fraction ($\delta\nu_c$)
and the critical resistance ($\rho^*$)
are related through $\rho^* (T)  \!=\!  e^{{\delta\nu_c}\over{\nu_0}}$. This
shift 
is next to be compared to the difference ($\delta\nu_c$)  as it is obtained
from the definitions 
$\sigma_{xy}  \!=\!  1/2$ and 
${\rm d}\sigma_{xx} / {\rm d}B  \!=\! 0$. In the lower
inset of Fig. 2 we plot 
the $\delta\nu_c /\nu_c$ or, equivalently, $\delta B_c /B_c$ with varying $T$
for both cases. The two 
effects are comparable. 

Notice that the uncertainty $\delta\nu_c /\nu_c$ in the definition of
$\nu_c$ clearly shows the effect of macroscopic inhomogeneities (in
electron density)
which cause $\nu_c$ to be slightly different in the different regions of the sample where
the $\rho_{xx}$ and
$\rho_{xy}$ are being probed. Fig. 2 therefore indicates that the weak or
{\em marginal} 
$T$ dependence of $\sigma_{xx}^*$ is, in fact, an {\em inhomogeneity} effect. 
This lack of universality in $\sigma_{xx}^*$ also shows up in the different
data sets taken at
different experimental runs. After heating the system up to room $T$ and
then cooling down again
one usually finds that the $B_c$ has shifted (indicating that the electron
density has changed)
along with a shift in $\sigma_{xx}^*$ (indicating that the inhomogeneity
profile of the density has changed).

In summary we can say that the $PP$ and $PI$ transitions are the same.
This is in complete agreement 
with the predictions of the renormalization theory.$^2$ 
We have shown that the critical conductance $\sigma_{xx}^*$ 
as well as the exponent of the $PI$ transition are weakly 
affected by the (weak) macroscopic inhomogeneities in the sample. Our data
retain fundamental
aspects such as the electron-hole symmetry in the $\sigma_{xx}
,\sigma_{xy}$ diagram. It is important that
this symmetry is not confused with the statement of duality$^{16}$ which
is, in fact, not verified by
our experiments. 

The marginal $T$ dependence in $\sigma_{xx}^*$ is common to both the $PP$
and $PI$ transitions in our samples.
This was previously also observed.$^1$
It is important to note that Eq. 3, upon modification, is applicable
to the $PP$ transitions as well. For example, for the
$2\!\!\rightarrow\!\!1$ transition  
Eq. 3 is modified according to 
$\kappa'  \!-\! \kappa  \!=\!  
{{\rm d}\ln (\sigma_{xx}^{*2} + 9/4)\over{\rm d}\ln T}$. 
By inserting the $\sigma_{xx}^*$ data
we find in this case $\kappa'  \!-\! \kappa  \!<\!  .01$ which is well within the experimental error.
This result explains why a single 
exponent $\kappa  \!=\! \kappa'  \!=\! .42 \pm .04$ was previously extracted from the $2\!\!\rightarrow\!\!1$ 
transition as well as from the
higher Landau level experimental data over a wide range in $T$ ($4K \!-\! 20mK$).
By combining the results for the $PP$ and $PI$ transitions we conclude that
$\kappa  \!=\!  .42$ stands for the {\em universal} critical exponent of the quantum phase 
transition. The numerical value $\kappa'  \!=\!  .55$ on the other hand is the result of macroscopic
inhomogeneities. Following Eq. 3 it represents an {\em effective} exponent.
A deeper understanding of the inhomogeneity effects
demands a commitment to research
on specifically grown $InGaAs/InP$ heterostructures.

This research was supported in part by the {\em Dutch Science Foundation FOM}
and by the {\it National Science Foundation} under Grant N$^{\rm os}$ 
{\it PHY94-07194} and {\it DMR-9311091}.

\begin{figure}
\begin{center}
\setlength{\unitlength}{1mm}
\begin{picture}(80,70)(0,0)
\put(0,0)
{\epsfxsize=7cm{\epsffile{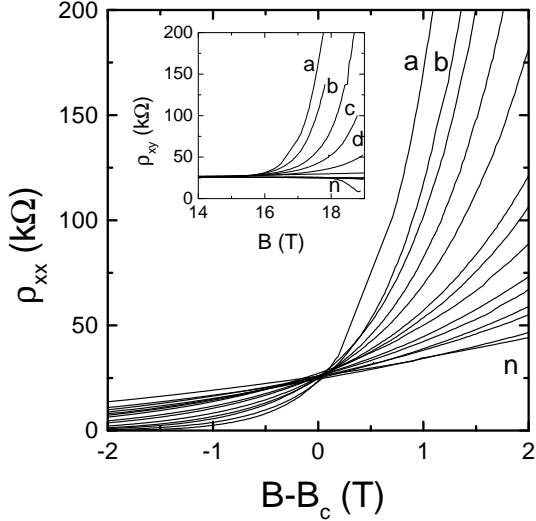}}}
\end{picture}
\caption{$\rho_{xx}$ and $\rho_{xy}$ data with varying $B$. The curves are labelled $a,b,\ldots,n$
and the corresponding $T$'s are 0.13, 0.21, 0.26, 0.35, 0.47, 0.59,
0.83, 1.04,
1.4, 1.5, 1.9, 2.2, 3.1 and 4.2K. $B_c$ is the critical $B$ (see text).} 
\label{Fig. 1}
\end{center}
\end{figure}

\begin{figure}
\begin{center}
\setlength{\unitlength}{1mm}
\begin{picture}(80,70)(0,0)
\put(0,0)
{\epsfxsize=7cm{\epsffile{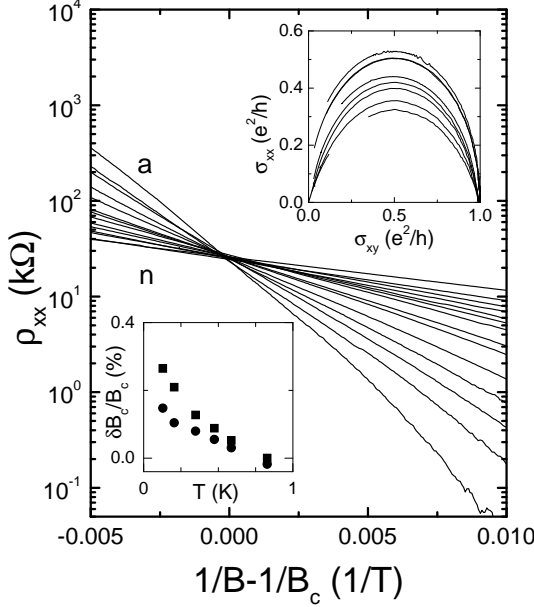}}}
\end{picture}
\caption{$\rho_{xx}$ data on a log scale with varying $1/B$.
The labels and $T$'s are the same as in Fig. 1.
Upper inset: $\sigma_{xx}$ vs. $\sigma_{xy}$ at various $T$ in the range
$1K-130mK$. Lower inset:
$\delta B_c / B_c$ vs. $T$. The squares are the data derived from $\rho^*$, 
Eqs 2 and 4, the circles are the data obtained from the different
definitions of $B_c$ (see text).} 
\label{Fig. 2}
\end{center}
\end{figure}
 
\begin{figure}
\begin{center}
\setlength{\unitlength}{1mm}
\begin{picture}(80,70)(0,0)
\put(0,0)
{\epsfxsize=7cm{\epsffile{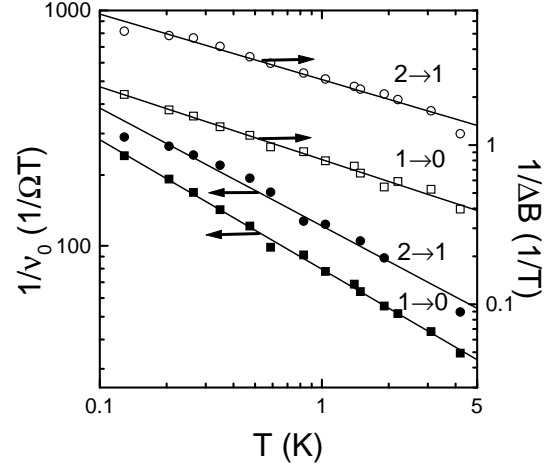}}}
\end{picture}
\caption{Left axis: $1/\nu_0$ vs. $T$ for the 
$2\!\rightarrow\! 1$ plateau transition 
(full circles, $\kappa' \!=\! 0.51$) and the $PI$ transition (full squares, 
$\kappa' \!=\! 0.55$).
Right axis: $1/\Delta B$ vs. $T$ for the 
$2\!\rightarrow\! 1$ plateau transition 
(open circles, $\kappa \!=\! 0.42$) and the $PI$ transition (open
squares, $\kappa \!=\! 0.46$).} 
\label{Fig. 3}
\end{center}
\end{figure}

\begin{figure}
\begin{center}
\setlength{\unitlength}{1mm}
\begin{picture}(80,70)(0,0)
\put(0,0)
{\epsfxsize=7cm{\epsffile{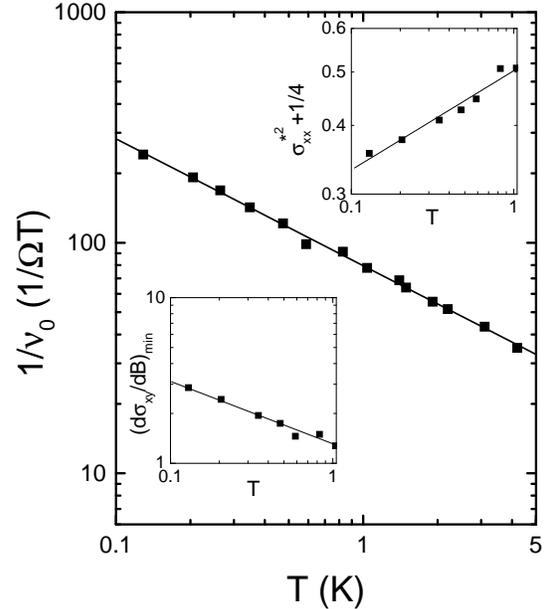}}}
\end{picture}
\caption{$1/\nu_0$ vs. $T$ for the $PI$ transition (full squares, 
$\kappa' \!=\! 0.55$). 
Upper inset: $\sigma_{xx}^{*2} \!+\! 1/4$ vs. $T$. 
The slope of the 
straight line equals $0.15$. Lower inset:
$({{\rm d}\sigma_{xy}\over {\rm d}B})_{\rm min}$ vs. $T$. 
The slope of the straight line equals $0.43$.}
\label{Fig. 4}
\end{center}
\end{figure}

\end{multicols}

\end{document}